\documentclass{article}
\usepackage[utf8]{inputenc}
\pdfoutput=1
\usepackage{graphicx}
\usepackage{graphbox}
\usepackage{xspace}

\usepackage{hyperref}
\hypersetup{colorlinks,allcolors=black}

\usepackage{amsmath}
\usepackage{lipsum}
\usepackage{verbatim}
\usepackage{mathtools}

\usepackage{subfig}
\usepackage{siunitx}

\usepackage{lineno}  

\usepackage[
top    = 2.5cm,
bottom = 2.5cm,
left   = 2.5cm,
right  = 2.5cm]{geometry}

\usepackage[section,below]{placeins}  
\usepackage[nottoc,numbib]{tocbibind}

\begin{document}

%



\begin{titlepage}
%
\vspace*{-1.5cm}
\centerline{\large Snowmass 2021 (DPF Community Planning Exercise) }
\vspace*{0.5cm}
\noindent
\begin{tabular*}{\linewidth}{lc@{\extracolsep{\fill}}r@{\extracolsep{0pt}}}
\vspace*{-1.5cm}\mbox{\!\!\!\includegraphics[width=.14\textwidth]{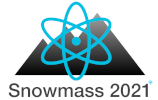}} & &%
\\
 & & \\   
 & & May 5, 2022 \\ 
 & & 
\end{tabular*}

\vspace*{0.2cm}

\begin{center}
\huge \textbf{Review of CMS contribution \\ \vspace{-0.3cm} to Hadron Spectroscopy} \\
\Large \textbf{(White Paper)}
\end{center}
\vspace{0.2cm}

\begin{center}
\large
Ruslan Chistov$^1$, Vaia Papadimitriou$^2$, Sergey Polikarpov$^1$, \\ Alexis Pompili$^3$, Alberto Sanchez-Hernandez$^4$

\vspace{0.8cm}
\noindent
\normalsize
$^1$National Research Nuclear University “MEPhI”, Moscow, Russia \\
$^2$Fermi National Accelerator Laboratory, Batavia IL, U.S.A. \\
$^3$Dipartimento Interateneo di Fisica, Università degli Studi di Bari, Bari, Italy \\
$^4$Physics Department, CINVESTAV, Mexico City, Mexico
\end{center}

\vspace{0.8cm}
\begin{abstract}
\linespread{1.2}
\noindent
\normalsize
In this brief White Paper we glance over the studies carried out so far by the CMS Collaboration in the field of conventional and exotic hadron spectroscopy. We highlight the most relevant achievements and discuss the future perspectives of this engagement.
\end{abstract}
\vspace{0.8cm}
\normalsize
\tableofcontents
\end{titlepage}


\clearpage
\pagestyle{plain}
\pagenumbering{arabic}
\setcounter{page}{2}

\section{Introduction}

The Hadron Spectroscopy has been experiencing a renaissance in the last 20 years thanks to the experimental findings at the B-factories, the Tevatron and the LHC. 
Many of the - more than 50 - experimental discoveries of both conventional and exotic QCD states have been performed in the LHC era, most of them by LHCb but also the CMS experiment has provided relevant contributions. Exotic states (the so-called XYZ states) differ from conventional hadrons (mesons, as quark-antiquark bound states and baryons, as 3 quark bound states) 
since they must be characterized by a quark composition clearly exotic (with 4, 5 or even 6 quark configuration).
The development of several theoretical models has not provided yet a unified explanation for these states or at least for a good majority of them.
Even the nature of the first charmonium-like state discovered in 2003, the $X(3872)$ is still a puzzle.
While the exotic states discovered so far need to be understood within a possibly consistent framework, the ongoing and future experimental observations and measurements are crucial 
both in establishing them and in deriving their main features (quantum numbers, decay channels and production modes). Indeed, some of these states need a confirmation by a second experiment, the majority of them have been observed only in a single production mechanism and the quantum numbers of many of them are not well established.
The following brief review of the main CMS achievements and contributions in this sector clarifies and highlights the strength points of CMS in producing future results. Theory and other experiment references mentioned in this document are included in the CMS publications that make up the final bibliography.

\vspace{0.5cm}
\section{Conventional spectroscopy}

In the beauty meson sector CMS has been able to experimentally disentangle for the first time the lower radial excitations of the $B_c^{+}$ state, the $B_c^+(2S)$ and the $B_c^{*+}(2S)$~\cite{bc2s}, and to measure the mass of the former~\cite{bc2s} and their cross section ratios, namely both ratios relative to the $B_c^{+}$ production and the relative $B_c^{*+}(2S)$ to $B_c^+(2S)$ one~\cite{bc2s-xsec}. 
Moreover, by studying the P-wave $B_s^{0}$ meson states in the $B^{(*)+}K^-$ and $B^{(*)0}K_{s}^{0}$, CMS has observed for the first time the decay $B_{s2}^{*}(5840)^{0} \to B^{0}K_{s}^{0}$, together with the first evidence for the decay $B_{s1}(5830)^{0} \to B^{*0}K_{s}^{0}$~\cite{bs2star}, and measured precisely their masses as well as the mass difference between $B^{(*)0}$ and $B^{(*)+}$.

In the beauty baryon sector CMS has performed the first observations of: 1) the beauty strange 
$\Xi_{b}^{*0}$ (with $J^P = 3/2^{+}$) decaying to $\Xi_{b}^{-}\pi^{+}$~\cite{xib}, 
and 2) the lightest orbitally excited $\Xi_{b}^{**-}$ named as $\Xi_{b}(6100)^{-}$ decaying to $\Xi_{b}^{-}\pi^{+}\pi^{-}$~\cite{xibstst}. 
Moreover, CMS has confirmed the existence of four $\Lambda^{0}_{b}$ excited states 
[$\Lambda_{b}(5912)^0$, 
$\Lambda_{b}(5920)^0$, 
$\Lambda_{b}(6146)^0$ and
$\Lambda_{b}(6152)^0$, all decaying to $\Lambda^{0}_{b} \pi^{+}\pi^{-}$] 
and firstly hinted the existence of a further excited state later called $\Lambda_{b}(6072)^0$~\cite{lambdabstst}.

In the bottomonium sector CMS has been able to observe, for the first time, the resolved 
$\chi_{b1}(3P)$ and $\chi_{b2}(3P)$ states and to measure their masses in the $\Upsilon (3S)\gamma$ decay channel by reconstructing the low-energy photons by pair conversion in the silicon tracker material~\cite{chib}.

It is worthy to mention also the first observations of rather rare decays with $K^{0}_{s}$, $\Lambda$ and $\phi$ mesons in the final state, together with a charmonium exploited at trigger level: 
$B^+ \to \psi(2S) \phi K^{+}$~\cite{bpsiphik}, 
$\Lambda^{0}_{b} \to J/\psi \Lambda \phi$~\cite{lbjpsiLphi}, 
$B^{0}_{s} \to X(3872) \phi$~\cite{bsXphi}, 
$B^{0}_{s} \to \psi(2S) K^{0}_{s}$ and 
$B^{0} \to \psi(2S) K^{0}_{s} \pi^{+} \pi^{-}$~\cite{b0K0s}.

\vspace{0.5cm}
\section{Exotic spectroscopy}

CMS has devoted few studies to contribute to the understanding of the nature of the $X(3872)$. 
In the first study CMS has inclusively reconstructed the $X(3872)$ signal
in the $J/\psi \pi^{+} \pi^{-}$ final state~\cite{xprod}. 
The invariant mass distribution of the dipion system confirmed the correctness of the assumption of an intermediate two-body decay into $J/\psi \rho^{0}$, except for the approximation in the simulation.
The non-prompt fraction of the reconstructed $X(3872)$ candidates was measured and no dependence on the transverse momentum $p_{T}$ was observed. 
Finally the prompt production cross section as a function of $p_{T}$ was measured, thus complementing - at central rapidities - the LHCb forward measurement and confirming a result more than $3\sigma$ lower than the prediction for a pure S-wave molecular model by Artoisenet and Brateen.

Afterwords CMS has looked for the existence of the 'bottomonium counterpart' of the $X(3872)$, denoted as $X_{b}$, in the $\Upsilon (1S) \pi^{+} \pi^{-}$ mass spectrum till $11$~GeV~\cite{y1Spp}. 
No significant excess has been observed in the whole spectrum and in particular close to the open-beauty thresholds as suggested by the Swanson's molecular model and therefore an upper limit on the production cross section of the $X_{b}$ (times its branching fraction for the decay into the final state) has been provided.
Moreover, no excess is visible in the $\Upsilon (3S)\gamma$ invariant mass spectrum~\cite{chib}, thus excluding any effect induced by the $B\bar{B}^{(*)}$ thresholds that are nearby the $\chi_{b(1,2)}(3P)$ doublet.

The branching fraction (BF) measurement of the $B^{0}_{s} \to X(3872) \phi$ decay, observed for the first time~\cite{bsXphi}, confirms, in the beauty-strange system, the relevantly lower BF for the neutral B meson decays to $X(3872)$ plus a light meson with respect to that for the charged B meson, a feature not observed when considering the corresponding decays to $\psi(2S)$ plus a light meson; 
this circumstance has found an explanation in the tetraquark scenario by Maiani and collaborators.

CMS has also found the first evidence of inclusive $X(3872)$ production in Heavy Ions (HI) collisions~\cite{xPbPb}. 
The ratio of efficiency-corrected yields of prompt $X(3872)$ to prompt $\psi(2S)$ 
(times their BFs to $J/\psi \pi^{+} \pi^{-}$) hints, within the limited statistics available, that the $X(3872)$ is less suppressed than $\psi(2S)$ in PbPb collisions with respect to pp ones.

Four further exotic searches have been published by CMS, well beyond studying the $X(3872)$ and possible partners.
Exploiting only part of the Run 2, CMS has measured the $\Upsilon(1S)$ pair production cross section at $13$~TeV and used this process as standard reference in a search for narrow resonances decaying 
to $\Upsilon(1S) \mu^+ \mu^-$ (same final state and similar set of selection criteria), in an extended mass window from $16.5$~GeV to $27$~GeV~\cite{ymm}. 
No significant excess consistent with an heavy bottom tetraquark signal has been observed above the background expectation; thus an upper limit on the product of the production cross section of a $b\bar{b}b\bar{b}$ state and the BF to a final state of four muons via an intermediate $\Upsilon(1S)$ state has been given in the mass window between $17.5$~GeV and $19.5$~GeV (that is around 4 times the mass of the bottom quark).

Furthermore, by reconstructing a large sample of $B^+ \to J/\psi \phi K^{+}$ decays, CMS confirmed~\cite{y4140} the peaking structure (denoted as $X(4140)$) close to the kinematic threshold of the $J/\psi \phi$ mass spectrum, observed by CDF with very low statistics. CMS hinted also a second peaking structure compatible with the $X(4274)$ state whose evidence was reported earlier by CDF. This result triggered, later, LHCb to perform a full amplitude analysis and observe six states in the $J/\psi \phi$ mass spectrum.

By studying the $B^{+} \to J/\psi \bar{\Lambda} p$ decay mode, CMS has found~\cite{modindep} the invariant mass distributions $J/\psi \bar{\Lambda}$, $J/\psi p$ and $\bar{\Lambda} p$ to be inconsistent with the pure phase space hypothesis. By using a model-independent angular analysis, the observed invariant mass distributions have turned out to be consistent with the contributions from excited kaons decaying to the $\bar{\Lambda} p$ system.

Finally, CMS has searched~\cite{x5568} for the hypothetical $X(5568)$ state claimed by D0, reporting a negative result
and thus complementing a similar result from LHCb in a central kinematic region (similar to that of the D0 experiment). The Upper limits on the relative production rate of the $X(5568)$ state and the $B^0_s$ meson, 
times the unknown BF of $X(5568)$ to $B^0_s \pi^{\pm}$ are the most stringent to date and have been provided for different hypothetical values of the natural width.

\vspace{0.5cm}
\section{Future perspectives}

The CMS Collaboration is clearly engaged in performing searches and measurements in the field of conventional and exotic spectroscopy. 
This commitment is longstanding (since the Run 1) and has clearly profited from the higher statistics collected in the Run 2. 
In some of the above searches or measurements the full Run 2 data set has to be fully exploited yet. 
Many analyses have been carried out but many others are ongoing or in the to-do list. 
Many analyses are prohibited by huge backgrounds, trigger constraints or reconstruction or identification limitations, 
however many others can be carried out with competitive results by exploiting some excellent features of the detector and reconstruction algorithms.

The data that are going to be collected in Run 3 and Run 4 can certainly help to achieve very interesting new and updated results, integrating or complementing LHCb results. This will be possible by carefully designing the triggers for the future data taking campaigns characterized by harsher experimental conditions. The availability of tracking information at the Level-1 trigger will be crucial to retain the full physics potential when pile up conditions expected in Phase 2 ($<$PU$>\sim 140-200$) will be realized.
Moreover the new timing layer will allow in Phase 2 not only some hadronic PID capabilities for the softer tracks but also an upgrade of the 3D vertex fit to a 4D one thus allowing precision timing for charged hadrons and converted photons and an effective pile up mitigation.

Particular strengths of the CMS detector and reconstruction software, for the spectroscopy studies, are: 
\begin{itemize}
    \item the good efficiency for the low-momentum tracks both prompt and displaced from the Primary Vertex. The latter are crucial for the reconstruction of the $K^{0}_{s} \to \pi \pi$ and self flavour-tagging $\Lambda \to p \pi$ decays; these decays are very important for the reconstruction of several baryonic decay chains.
    \item The precise and competitive photon conversions for the radiative spectroscopic transitions with photon energies larger than $400$~MeV.
    \item The large muons' acceptance which is useful in particular for bottomonium signals extraction  and their associated production with charmonia and bosons, for double bottomonia production and, in general, for multi-muon final states.
\end{itemize}

By lacking a hadronic particle identification, CMS is clearly more competitive when dealing with signatures with $K^{0}_{s}$, $\Lambda$ and $\phi$ reconstructed mesons that allow to fight the overwhelming backgrounds associated to huge track multiplicity in the event. Very rare radiative decays can be searched for by exploring the usage of calorimeter photons when resolution is not crucial.

In general, beauty hadrons and quarkonia production measurements will be provided in a phase-space complementary to LHCb and ALICE (in HI collisions) and it is worthy to mention that CMS would be capable to search for QCD exotics in HI collisions ($X(3872)$ and beyond) that will be hardly doable at ALICE.

\bibliography{RF07_Snowmass21WP_CMS_HadSpectr}

\providecommand{\href}[2]{#2}\begingroup\raggedright\begin{thebibliography}{10}

\bibitem{bc2s}
{CMS Collaboration}, {\it {Observation of Two Excited $B^+_\mathrm{c}$ States
  and Measurement of the $B^+_\mathrm{c}$(2S) Mass in pp Collisions at
  $\sqrt{s} =$ 13 TeV}},  {\em Phys. Rev. Lett.} {\bf 122} (2019) 132001,
  [\href{http://arxiv.org/abs/1902.00571}{{\tt arXiv:1902.00571}}].

\bibitem{bc2s-xsec}
{CMS Collaboration}, {\it {Measurement of B$_\mathrm{c}$(2S)$^+$ and
  B$_\mathrm{c}^*$(2S)$^+$ cross section ratios in proton-proton collisions at
  $\sqrt{s} =$ 13 TeV}},  {\em Phys. Rev. D} {\bf 102} (2020) 092007,
  [\href{http://arxiv.org/abs/2008.08629}{{\tt arXiv:2008.08629}}].

\bibitem{bs2star}
{CMS Collaboration}, {\it {Studies of ${\mathrm {B}} ^{*}_{\mathrm
  {s}2}(5840)^0 $ and ${\mathrm {B}} _{{\mathrm {s}}1}(5830)^0 $ mesons
  including the observation of the ${\mathrm {B}} ^{*}_{\mathrm {s}2}(5840)^0
  \rightarrow {\mathrm {B}} ^0 \mathrm {K} ^0_{\mathrm {S}} $ decay in
  proton-proton collisions at $\sqrt{s}=8\,\text {TeV} $}},  {\em Eur. Phys. J.
  C} {\bf 78} (2018) 939, [\href{http://arxiv.org/abs/1809.03578}{{\tt
  arXiv:1809.03578}}].

\bibitem{xib}
{CMS Collaboration}, {\it {Observation of a new Xi(b) baryon}},  {\em Phys.
  Rev. Lett.} {\bf 108} (2012) 252002,
  [\href{http://arxiv.org/abs/1204.5955}{{\tt arXiv:1204.5955}}].

\bibitem{xibstst}
{CMS Collaboration}, {\it {Observation of a New Excited Beauty Strange Baryon
  Decaying to $\Xi^-_\mathrm{b} \pi^+ \pi^-$}},  {\em Phys. Rev. Lett.} {\bf
  126} (2021) 252003, [\href{http://arxiv.org/abs/2102.04524}{{\tt
  arXiv:2102.04524}}].

\bibitem{lambdabstst}
{CMS Collaboration}, {\it {Study of excited $\Lambda_{\mathrm{b}}^0$ states
  decaying to $\Lambda_\mathrm{b}^0\pi^+\pi^-$ in proton-proton collisions at
  $\sqrt{s}=$ 13 TeV}},  {\em Phys. Lett. B} {\bf 803} (2020) 135345,
  [\href{http://arxiv.org/abs/2001.06533}{{\tt arXiv:2001.06533}}].

\bibitem{chib}
{CMS Collaboration}, {\it {Observation of the $\chi_\mathrm{b1}(3P)$ and
  $\chi_\mathrm{b2}(3P)$ and measurement of their masses}},  {\em Phys. Rev.
  Lett.} {\bf 121} (2018) 092002, [\href{http://arxiv.org/abs/1805.11192}{{\tt
  arXiv:1805.11192}}].

\bibitem{bpsiphik}
{CMS Collaboration}, {\it {Observation of the decay $B^+ \to \psi(2S)
  \phi(1020) K^+$ in pp collisions at $\sqrt s =$ 8 TeV}},  {\em Phys. Lett. B}
  {\bf 764} (2017) 66, [\href{http://arxiv.org/abs/1607.02638}{{\tt
  arXiv:1607.02638}}].

\bibitem{lbjpsiLphi}
{CMS Collaboration}, {\it {Observation of the $\Lambda_\mathrm{b}^0 \to$
  J/$\psi \Lambda \phi$ decay in proton-proton collisions at $\sqrt{s}=$ 13
  TeV}},  {\em Phys. Lett. B} {\bf 802} (2020) 135203,
  [\href{http://arxiv.org/abs/1911.03789}{{\tt arXiv:1911.03789}}].

\bibitem{bsXphi}
{CMS Collaboration}, {\it {Observation of the B$^0_\mathrm{s}\to $X(3872)$\phi$
  decay}},  {\em Phys. Rev. Lett.} {\bf 125} (2020) 152001,
  [\href{http://arxiv.org/abs/2005.04764}{{\tt arXiv:2005.04764}}].

\bibitem{b0K0s}
{CMS Collaboration}, {\it {Observation of
  B$^0$$\to$$\psi$(2S)K$^0_\mathrm{S}\pi^+\pi^-$ and
  B$^0_\mathrm{S}$$\to$$\psi$(2S)K$^0_\mathrm{S}$ decays}},  {\em Submitted to
  Eur. Phys. J. C} (2022) [\href{http://arxiv.org/abs/2201.09131}{{\tt
  arXiv:2201.09131}}].

\bibitem{xprod}
{CMS Collaboration}, {\it {Measurement of the $X$(3872) Production Cross
  Section Via Decays to $J/\psi \pi^+ \pi^-$ in $pp$ collisions at $\sqrt{s}$ =
  7 TeV}},  {\em JHEP} {\bf 04} (2013) 154,
  [\href{http://arxiv.org/abs/1302.3968}{{\tt arXiv:1302.3968}}].

\bibitem{y1Spp}
{CMS Collaboration}, {\it {Search for a New Bottomonium State Decaying to
  $\Upsilon(1S)\pi^+\pi^-$ in pp Collisions at $\sqrt{s}$ = 8 TeV}},  {\em
  Phys. Lett. B} {\bf 727} (2013) 57,
  [\href{http://arxiv.org/abs/1309.0250}{{\tt arXiv:1309.0250}}].

\bibitem{xPbPb}
{CMS Collaboration}, {\it {Evidence for X(3872) in Pb-Pb Collisions and Studies
  of its Prompt Production at $\sqrt {s_{NN}}$=5.02\,\,TeV}},  {\em Phys. Rev.
  Lett.} {\bf 128} (2022) 032001, [\href{http://arxiv.org/abs/2102.13048}{{\tt
  arXiv:2102.13048}}].

\bibitem{ymm}
{CMS Collaboration}, {\it {Measurement of the $\Upsilon$(1S) pair production
  cross section and search for resonances decaying to
  $\Upsilon$(1S)$\mu^+\mu^-$ in proton-proton collisions at $\sqrt{s} =$ 13
  TeV}},  {\em Phys. Lett. B} {\bf 808} (2020) 135578,
  [\href{http://arxiv.org/abs/2002.06393}{{\tt arXiv:2002.06393}}].

\bibitem{y4140}
{CMS Collaboration}, {\it {Observation of a Peaking Structure in the $J/\psi
  \phi$ Mass Spectrum from $B^{\pm} \to J/\psi \phi K^{\pm}$ Decays}},  {\em
  Phys. Lett. B} {\bf 734} (2014) 261,
  [\href{http://arxiv.org/abs/1309.6920}{{\tt arXiv:1309.6920}}].

\bibitem{modindep}
C.~Collaboration, {\it {Study of the $ {\mathrm{B}}^{+}\to \mathrm{J}/\psi
  \overline{\Lambda}\mathrm{p} $ decay in proton-proton collisions at $
  \sqrt{s} $ = 8 TeV}},  {\em JHEP} {\bf 12} (2019) 100,
  [\href{http://arxiv.org/abs/1907.05461}{{\tt arXiv:1907.05461}}].

\bibitem{x5568}
{CMS Collaboration}, {\it {Search for the X(5568) state decaying into
  $\mathrm{B}^{0}_{\mathrm{s}}\pi^{\pm}$ in proton-proton collisions at
  $\sqrt{s} = $ 8 TeV}},  {\em Phys. Rev. Lett.} {\bf 120} (2018) 202005,
  [\href{http://arxiv.org/abs/1712.06144}{{\tt arXiv:1712.06144}}].

\end{thebibliography}\endgroup


\end{document}